# A foretaste of Qatar 2022: decreased playing time of internationals after the Africa Cup of Nations

Otto Kolbinger[1], Chenyuyan Yang[1], Martin Lames[1]

[1]*Chair of Performance Analysis and Sport Informatics, Technical University of Munich, Munich, Germany*

Corresponding Author: Dr. Otto Kolbinger, Chair of Performance Analysis and sport Informatics, Technical University of Munich, Georg-Brauchle-Ring 60/62, 80992 München, Germany, Tel. +49.(0)89.289.24502, E-mail: otto.kolbinger@tum.de

**Abstract**

Due to the unfavourable climatic conditions in Qatar during summertime, the FIFA World Cup 2022 will be played during on-going seasons of the major European leagues. This study investigates how national teams' tournaments scheduled at such a time window impact the playing time of released players, using data from the Africa Cups of Nations (AFCON). For 262 internationals playing at the 2013, 2015 and 2021 AFCON, we compared the share of possible games and minutes played before and after the tournament using Mann-Whitney-U tests. We found a significant decrease of 3.3% for games (p=.029, $CL_{EffectSize}$=44.5%) and 3.1% for minutes played respectively (p=.018, $CL_{EffectSize}$=44.9%). For a subsample of 163 players, which played for the same club the preceding seasons, we found that these players tend to have played more in the second half of the previous season, resulting in a net decrease of 6.8% for games (p=.011, $CL_{EffectSize}$=42.3%) and 7.1% for minutes played (p=.007, $CL_{EffectSize}$=41.9%). Conclusions for the FIFA World Cup 2022 should only be drawn carefully as the number of released players was comparatively low. However, the findings give some indication that releasing clubs might suffer the rest of the season after this tournament.

**Keywords: player releases, internationals, FIFA World Cup 2022, Qatar**

**Introduction**

Selecting Qatar as host country for the 2022 FIFA World Cup was met with a lot of legitimate refusal due to political and environmental reasons. For example, people raised concerns about violations of human rights in the forefront of the tournament and were, unfortunately, proven to be right (Amnesty International, 2018; Ewers et al., 2020; Regueiro, 2020). Concerning the competition itself, sport medical scientists pointed at the extreme climatic conditions, which can lead to serious health issues for the performing athletes (Brocherie et al., 2013). One measure taken to tackle this issue was the decision to move the tournament from its traditional time frame (June or July) to the end of the year. A decision that came along with a potentially disadvantageous side effect for clubs of the top European leagues. For the first time, the clubs have to release their players during an on-going season.

Of course, the seasons of the big European leagues will be paused during the 2022 World Cup, including a defined period before and after the tournament. However, the overall break is shorter than the offseason between two seasons. As an example, the English Premier League will pause from November 13th to December 26th. The offseason embedding the last world cup lasted from May 13th to August 10th, 2018, and the one from the second to last world cup even from May 11th to August 16th, 2014. Roughly spoken, the break including the major tournament got reduced by more than half.

A study from Lames and Kolbinger (2011) about the effects of the FIFA World Cup 2010 on club's performance provided indications that teams in the German Bundesliga already suffered from releasing players to major tournaments in-between seasons. An effect that could also be found in the English Premier League, the Italian Serie A and the French Ligue 1 (Kolbinger & Lames, 2012). In more detail, for both studies, the authors found that the link between the number of internationals in a team and the final ranking changed after the FIFA World Cup 2010. For the 2009/10 season, the leagues showed a strong positive correlation between the number of released players and the number of points won. This correlation decreased significantly and with a strong effect size, and even turned around for one league (Bundesliga), at the beginning of the 2010/11 season. A pattern the authors ascribed to the shortened preparation period and the lack of recovery for the released player, which leads to a decline of team's performance for such clubs that did send comparatively many players to the world cup.

Longley (2012) later framed this phenomenon as "fatigue theory", describing a similar effect for NHL teams that had to release players to the Olympics from 1998 to 2010. He also used a simple approach by measuring the change of the goal-differential of a team in relation to the number of released players. Cairney et al. (2015) extended this approach by deploying a growth curve model approach in which they controlled for further factors like previous team success. They found the same effect for the 1998, 2002, 2006 and 2014 Olympic games, however, stating that the respective performance differences were rather small. Bremer and Cairney (2020) later looked whether the number of minutes played at the Olympics influences those performance differences, but found no such effect. Throughout the paper, the authors point at various instances of players suffering season-ending injuries during the Olympics, like John Tavares or Aleksander Barkov at Sochi 2014. However, these players were not part of their analysis, as Bremer and Cairney (2020) used points per game as performance indicator.

Not including player's availability might be a substantial shortcoming. In football, findings of Hägglund et al. (2013), among others, indicate that the number of injured players is negatively affecting team success. In more detail, teams with more injuries, or vice versa: lesser player's availability, on average gained less points and, therefore, achieved worse rankings. Nevertheless, we are aware of no studies that have investigated, if players are more likely to miss games after major tournaments. In particular, after major tournaments that take place during an on-going season.

This study wants to target this shortcoming by investigating the change of playing time of players participating in three Africa Cup of Nations (hereinafter AFCON) tournaments. As the 2022 FIFA World Cup, this tournament is traditionally played during an on-going season and might provide indications if and to which extent clubs have to expect decreased availability of released players after Qatar 2022. Thus, AFCON data from the past is used for quasi-experimentally simulating the effects for the FIFA World Cup 2022 played in November and December 2022. In more detail, the aim of this study is to investigate the influence of the AFCON on games and minutes played after the tournaments for players of the top five European leagues.

**Materials and methods**

*Sample*

For our analysis, we retrieved data from the last three AFCONs played at the beginning of a calendar year and, therefore, during on-going European competitions: 2015, 2017 and 2021. Note that the AFCON 2019 was played in-between two seasons and the AFCON 2021 was postponed to 2022. As described in the introductory section, we limited our study to AFCON participants playing in one of the big five European leagues, as these leagues released the most players to the last world cup in 2018. To be eligible for our analysis, a player had to play for the same club for the majority of the first and second half of the season in which the AFCON took place. Thus, players who joined their club late in the pre-season transfer window were included, but not such players who joined or left their club in the transfer mid-season window.

Overall, 262 players (*Complete_Sample*) fulfilled these criteria for the three investigated AFCONs ($n_{2015} = 81$, $n_{2017} = 65$, $n_{2021} = 116$). For each of those players we retrieved game-by-game playing time for all competitions his club participated from www.transfermarkt.de. We chose to use this website as our data source, as it is known to provide reliable data not just for playing time but also for additional information for games in which players did not play, like present injuries or suspensions (Leventer et al., 2018). For all players of our sample, we further checked if they played for the same club in the season before the AFCON as well. This applied to 163 players (*Sample_With_Previous_Season*) for which we collected the game-by-game playing time for the previous season as well.

Analysis

The aim of this study is to explore the influence of participation at the AFCON on player availability, which we operationalized as games and minutes played. Depending on a team's league and progress in further national and international competitions, like the UEFA Champions league, the number of possible games and, consequently, minutes can vary. This is why for each player and half-season, we calculated the share of possible games played and minutes played respectively. As an example, Eric Maxim Choupo-Moting's club Bayern Munich's schedule in the first half of the 2021/22 season consisted of 17 league games (Bundesliga), 3 domestic cup games (DFB-Pokal and Supercup) and 6 UEFA Champions League games. Eric Maxim Choupo-Moting appeared in 46.2% (12) of those 26 games and played for 10.6% (248) of the possible 2340 minutes. Note that due to the data structure used by www.transfermarkt.de we set the maximum playing time as 90 minutes per game and 120 minutes for games going into overtime. Games (and minutes) played during the period in which players were released to the AFCON were not included in the number of possible games. Thus, a player who appeared in all games after an AFCON participated in 100% of the possible games. Also, games in which a player got suspended were not included in the number of possible games.

We ran two different kinds of analysis with our data. First, for the season in which an AFCON took place, we compared the shares of possible games and minutes played before the tournament (*Pre_AFCON*) to the shares after it (*Post_AFCON*). Second, for all players that also played in the same season the year before the AFCON, we compared the difference found for the AFCON season (*Diff_AFCON*) to the difference found for the previous season (*Diff_PrevSeas*). For such previous seasons, in which no AFCON took place, the difference was calculated for the first (*PrevSeas_First*) and the second (*PrevSeas_Second*) half of the season, as the AFCON usually is played at the beginning of the year. As neither the share of games nor the share of minutes nor the respective differences were normally distributed, we compared the groups using Mann-Whitney-U tests. For each comparison we also report the so-called common language effect size ($CL_{EffectSize}$), which has originally been introduced by McGraw and Wong (1992). To illustrate $CL_{EffectSize}$ for one of our measures, it can be interpreted as the probability that a randomly selected share of minutes played by one player before the AFCON (*Pre_AFCON*) is higher than a randomly selected share of minutes by one player PostAFCON.

# Results

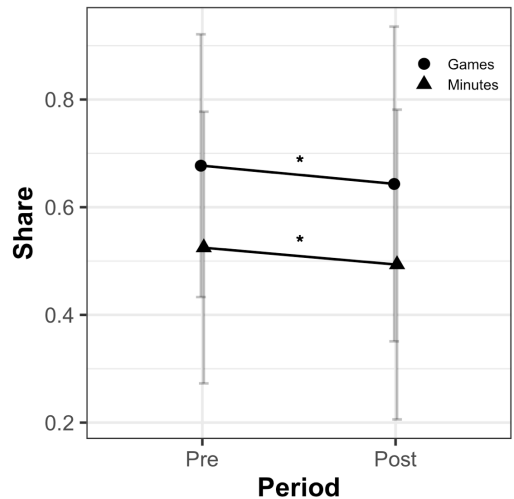

*Figure 1. Mean and standard deviations of the share of possible games (circles) and minutes (triangles) played in the PreAFCON and the PostAFCON period. Points and error bars are slightly shifted to improve recognizability. * mark significant differences on a .05 level.*

Figure 1 illustrates the share of games and minutes played before and after the AFCON tournaments. The 262 players of the *Complete_Sample* appeared on average in 67.7% (SD = 24.4%) of the possible games in a season before being released to the AFCON. This value significantly decreased to 64.3% (SD = 29.2%) for the remainder of the season after the tournament (Z = 0.72, p = .029), equalling a common language effect size of 44.5% that players appeared in more games in the *Pre_AFCON* period. The same trend can be observed for the share of minutes played, which decreased from 52.5% to 49.4% ($SD_{Pre\_AFCON}$ = 25.2%; $SD_{Pre\_AFCON}$ = 28.8%; $CL_{EffectSize}$ = 44.9%, Z = 1.20, p = .018). The variance of the playing time was significantly higher in terms of possible games as well as minutes played after the AFCON (*Games*: F = 0.70, p = .004; *Minutes*: F = 0.77, p = 0.035).

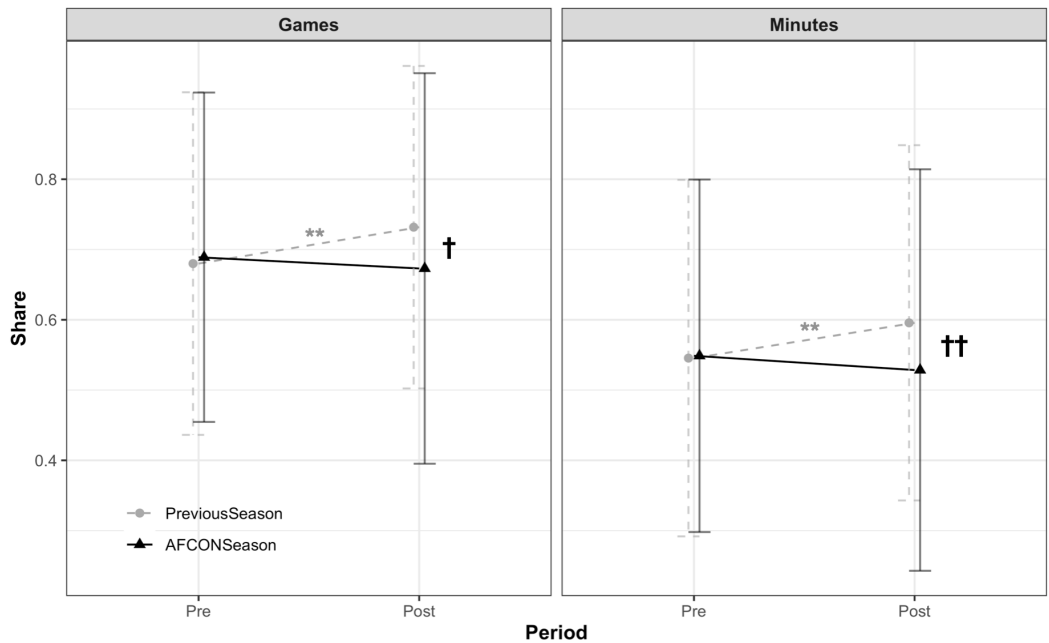

*Figure 2. Development of the mean and standard deviations of the share of possible games and minutes played in the season before the AFCON (PreviousSeason; grey) and the AFCONSeason (black). Points and error bars are slightly shifted to improve recognizability. \*\* mark significant differences on a .01 level between the two periods in a season . † and †† mark differences for the development between the two seasons on a .05 level and a .01 level respectively.*

The patterns for the subsample *Sample_With_Previous_Season* are displayed in figure 2. For players of this sample that also played for the same team in the previous season the pattern was different than above for these previous seasons. They played a higher share of possible games (73.2%) and minutes (59.6%) in the second half of the season, compared to the first half (*Games*: 68.0%; *Minutes*: 54.6%). A difference with a common language effect size of 59.6% (*Games*) and 59.4% (*Minutes*) respectively, which was also significant (*Games*: $Z = 2.12$, $p = .003$; *Minutes*: $Z = 1.83$, $p = .001$). For the previous season, the variances of both periods were equal (*Games*: $SD^2_{PrevSeas\_First} = 5.95\%$, $SD^2_{PrevSeas\_Second} = 5.24\%$, $F = 1.13$, $p = .442$; *Minutes*: $SD^2_{PrevSeas\_First} = 6.45\%$, $SD^2_{PrevSeas\_Second} = 6.40\%$, $F = 1.01$, $p = .962$).

For the AFCON season, this subsample showed the same pattern when comparing the periods before the AFCON and after the AFCON as the complete sample (*Games$_{Pre\_AFCON}$* = 68.9% vs. *Games$_{Post\_AFCON}$* = 67.3%, $Z = 0.09$, $p = .608$, $CL_{EffectSize}$ =

47.3%; $Minutes_{Pre\_AFCON}$ = 54.9% vs. $Minutes_{Post\_AFCON}$ = 52.9%, Z = 0.48, p = .389, $CL_{EffectSize}$ = 46.7%). If the differences for possible games played between both periods for the seasons are compared, the Mann-Whitney-U test revealed a significant difference with a CL effect size of 42.3% ($Games_{Diff\_PrevSeas}$ = -5.18, $Games_{Diff\_AFCON}$ = 1.60, Z = 2.43, p = .011). The same applied for minutes played, with a slightly stronger CL effect of 41.9% that players had a higher share for the *AFCONSeason* ($Minutes_{Diff\_PrevSeas}$ = -5.02, $Minutes_{Diff\_AFCON}$ = 2.03, Z = 2.70, p = .007).

**Discussion**

The results of this study provide an intriguing foretaste for the FIFA 2022 World Cup in Qatar. We could show that players, which were released to three of the last four Africa Cup of Nations, a tournament traditionally held during an on-going season, played significantly less after the tournament. The decrease of around 3% in an AFCON season becomes even more meaningful, considering that this cohort of players usually showed an increase of playing time in the second half of seasons without an AFCON. Thus, the net loss of playing time is around 7%, equalling almost one and a half games per half season per player (including international and domestic cup games). As top teams very often release a double-digit number of players to World Cups, for example Manchester City released a leading 16 players 2018, the net loss of player's availability per team can easily add up to over 20 games.

To have one player not available for one and a half games does probably not mean a direct financial disadvantage for a club. Players of the English Premier League earned on average 3.64 million US$) in the 2019/20 season (Kennedy & Kennedy, 2021), which equals around 80,000 US$ per game. For each day a player was released to FIFA World Cup 2018 – there were no numbers for the FIFA World Cup 2022 available when this manuscript was submitted – the releasing club received 8,530 US$

(FIFA, 2021). This amount was due for each day from two weeks before the tournament until the day after the last game of the respective team, resulting in a total between 255,900 US$ and 400,910 US$., depending on the progress of the national team during the tournament. It can be assumed that this number will be higher for the FIFA World Cup 2022 and, therefore, that there is no direct financial disadvantage for a releasing club, even if we consider that internationals on average get up to four-time higher salaries (Lucifora and Simmons, 2003). Of course, the cost-benefit analysis for releasing clubs is much more complicated, as the participation at such tournaments also affects the transfer values of players to name but one thing (see e.g., Gürtler et al., 2015).

As mentioned above, a negative impact on the trade-off for clubs could result from a decrease of performance of the released players. Findings from previous studies about such an effect in hockey at least indicate that players perform slightly worse after participating in major tournaments during an on-going season (Longley, 2012; Cairney et al., 2015; Bremer & Cairney, 2020). Further works from Kolbinger and Lames (2011) as well as Kolbinger and Lames (2012) respectively already showed that football teams of the Top 5 leagues that had to release a lot of players tend to perform worse after major tournaments. This impact is in particular meaningful as clubs send a varying number of players to such tournaments, ranging from 0 to 16 (Manchester City) at the last world cup.

Similar to the number of released players, the amount of time missed by those players is unevenly distributed over the clubs. Some players, like Abdul Rahman Baba in 2016/17 (at that time playing for Schalke 04), missed the complete remainder of the season after getting injured, whereas other players, like Ahmed Elmohamady in 2016/17 (Hull City), appeared in all possible games after the tournament. It can be assumed that injuries are a major reason for players missing games (Ekstrand et al. 2011) and, overall, the risk to suffer injuries is higher for players representing their countries at major tournaments. First, Waldén et al. (2007) found by investigating different European Championships that 11% of the players got injured during the tournament. Second, it is well-established that injury risk increases with the number of games played in a season (Carling et al., 2010; Leventer et al., 2019). As mentioned above, Häggelund et al. (2013) showed that performance is affected negatively by injuries.

Besides the potential decrease of performance and the increase of injury risk there is another disadvantage for releasing clubs. The findings of this study showed that the inter-individual differences in playing time were higher in the *PostAFCON* period, compared to the *PreAFCON* period. For the preceding season, the inter-individual differences were consistent for both halves of the season. Concluding carefully, this shows that releasing player during on-going seasons decreases the planning reliability regarding the roster composition for the remainder of the season after a major tournament.

In this study, we did only measure whether a player missed (or appeared in) a game, but not the reason for his absence, except if he was suspended. We argue that for our purpose it is not relevant if a player misses a game due to an acute injury, recovery reasons or a drop in performance. However, this is still a limitation of this study, as players can also not appear in games (or less) due to tactical reasons or changes in the clubs' roster to name but two more reasons. Further, even if injuries get announced by clubs, it's often not possible to map the injures to specific events.

Another limitation is that our data source transfermarkt.de caps the playing time per game at 90 minutes (and 120 minutes respectively), meaning that additional time gets neglected, which can vary between leagues (Prüßner & Siegle, 2015). This affects the possible and actual playing time in minutes. However, results for the share of minutes played were consistent with those of the share of games played.

**Conclusion**

Overall, this study showed that the playing time of players released to the Africa Cup of Nations decreases for the remainder of the season after the tournament. As a study from Perez (2021) already indicates that in particular teams with less resources perform worse during the AFCON, this will further fuel the debate about the way this tournament is currently scheduled. The 2023 tournament, originally scheduled to be played in-between two seasons, has been postponed to January 2024 (CAF, 2022).

      This year's FIFA World Cup 2022 will be played during an on-going season as well for the first time. Of course, conclusions for the FIFA World Cup 2022 should only be drawn carefully, as the number of released players to the AFCON was comparatively low and the setting differs from those of the AFCONs quite a lot (e.g., interrupted season, different time of the year). However, even if the respective seasons are interrupted, which at least avoids the negative affects during the tournament found by Perez (2021), clubs might be very concerned about the effects on their players. Previous research findings already showed a decrease of performance after such tournaments, due to a lack of recovery and shortened preparation time (e.g., Lames & Kolbinger, 2011; Longley, 2012). Findings from this study add to this by indicating a lower player's availability and less planning reliability for the clubs, which both affects a club's performance in a negative way as well. As a take-away from this study, one may expect a lower performance in the second leg of the season for clubs delegating many players to Qatar 2022, which could for example result in not qualifying for the UEFA Champions League. It will be interesting to see how clubs try to handle this situation

and how the FIFA tries to smoothen the trade-off between the clubs and the participating nations, e.g., by assigning more money to the releasing clubs.

Acknowledgments

The authors want to thank Sam Förster for his support in collecting data for the Africa Cup of Nations 2017.

References

Amnesty International (2018, September 26). Qatar: Migrant workers unpaid for months of work by company linked to World Cup host city. Amnesty International. https://www.amnesty.org/en/latest/news/2018/09/qatar-migrant-workers-unpaid-for-months-of-work-by-company-linked-to-world-cup/

Bremer, E., & Cairney, J. (2020). The Impact of Participation in the Olympics on Post-olympic Performance in Professional Ice Hockey Players. *Frontiers in Sports and Active Living, 2*:76. https://doi.org/10.3389/fspor.2020.00076

Brocherie, F., Girard, O., Farooq, A., & Millet, G. (2013). Influence of environmental temperature on home advantage in Qatari international soccer matches. In Peters, D., & O'Donoghue, P. (Eds.), *Performance analysis of sport IX* (pp. 65-70). Routledge.

CAF (2022, July 3). CAF President Dr Motsepe announces African Super League launch details, AFCON 2023 and Champions League key decisions. CAF Communications Department. https://www.cafonline.com/press-release/news/caf-president-dr-motsepe-announces-african-super-league-launch-details-afcon2023

Cairney, J., Joshi, D., Li, Y., & Kwan, M. (2015). The Impact of the Olympics on Regular Season Team Performance in the National Hockey League. *Journal of Athletic Enhancement, 4*(6), 1-7. https://doi.org/10.4172/2324- 9080.1000214

Carling, C., Orhant, E., & LeGall, F. (2010). Match injuries in professional soccer: inter-seasonal variation and effects of competition type, match congestion and positional role. *International journal of sports medicine, 31*(4), 271-276. https://doi.org/10.1055/s-0029-1243646

Ekstrand, J., Hägglund, M., & Waldén, M. (2011). Injury incidence and injury patterns in professional football: the UEFA injury study. *British journal of sports medicine, 45*(7), 553-558. http://dx.doi.org/10.1136/bjsm.2009.060582

Ewers, M. C., Diop, A., Le, K. T., & Bader, L. (2020). Migrant worker well-being and its determinants: The case of Qatar. *Social Indicators Research, 152*(1), 137-163. https://doi.org/10.1007/s11205-020-02427-3

FIFA (2021, June 15). FIFA World Cup club benefits programme. FIFA. https://digitalhub.fifa.com/m/63651247df9b8ba2/original/bxkl7wgkjygv0ar7scko-pdf.pdf

Gürtler, O., Lang, M., & Pawlowski, T. (2015). On the Release of Players to National teams. *Journal of Sports Economics, 16*(7), 695-713. https://doi.org/10.1177/1527002513503173

Hägglund, M., Waldén, M., Magnusson, H., Kristenson, K., Bengtsson, H., & Ekstrand, J. (2013). Injuries affect team performance negatively in professional football: an 11-year follow-up of the UEFA Champions League injury study. *British Journal of Sports Medicine, 47*(12), 738-742. http://dx.doi.org/10.1136/bjsports-2013-092215

Kennedy, D., & Kennedy, P. (2021). English premier league football clubs during the covid-19 pandemic: business as usual? *Soccer & Society, 22*(1-2), 27-34. https://doi.org/10.1080/14660970.2020.1797498

Kolbinger, O., & Lames, M. (2012). Europäische Spitzenvereine leiden unter WM-Abstellungen. In Jansen, C. T., Baumgart, C., Hoppe, M. W., & Freiwald, J. (Eds.), *Trainingswissenschaftliche, geschlechtsspezifische und medizinische Aspekte des Hochleistungsfußballs: 23. Jahrestagung der dvs-Kommission Fußball vom 24. - 26. November 2011 in Hannover* (pp. 39-44). Czwalina.

Lames, M., & Kolbinger, O. (2011). German Bundesliga Club's rankings in season 2010/11 are significantly affected by the number of players released to World Cup 2010. *International Journal of Performance Analysis in Sport, 11*(2), 308–313. https://doi.org/10.1080/24748668.2011.11868550

Leventer, L., Eek, F., & Lames, M. (2019). Intra-seasonal variation of injury patterns among German Bundesliga soccer players. *Journal of science and medicine in sport, 22*(6), 661-666. https://doi.org/10.1016/j.jsams.2018.12.001

Longley, N. (2012). The Impact of International Competitions on Competitive Balance in Domestic Leagues: The Case of the National Hockey League's Participation in the Winter Olympics. *International Journal of Sport Finance, 7*(3), 249-261.

Lucifora, C., & Simmons, R. (2003). Superstar effects in sport: Evidence from Italian soccer. *Journal of Sports Economics, 4*(1), 35-55. https://doi.org/10.1177%2F1527002502239657

McGraw, K. O., & Wong, S. P. (1992). A common language effect size statistic. *Psychological bulletin, 111*(2), 361. https://doi.org/10.1037/0033-2909.111.2.361

Pérez, L. (2021). Will We Lose If We Lose You? Players' Absence, Teams' Performance and the Overlapping of Competitions. *Journal of Sports Economics, 22*(6), 722-734. https://doi.org/10.1177/2F15270025211008499

Prüßner, R., & Siegle, M. (2015). Additional time in soccer–influence of league and referee. *International Journal of Performance Analysis in Sport, 15*(2), 551-559. https://doi.org/10.1080/24748668.2015.11868813

Regueiro, R. (2020). Shared Responsibility and Human Rights Abuse: The 2022 World Cup in Qatar. *Tilburg Law Review, 25*(1), 27–39. http://doi.org/10.5334/tilr.191

Waldén, M., Hägglund, M., & Ekstrand, J. (2007). Football injuries during European championships 2004–2005. *Knee Surgery, Sports Traumatology, Arthroscopy, 15*(9), 1155-1162. https://doi.org/10.1007/s00167-007-0290-3